%% file: eprint.tex
\documentclass[12pt]{article}
\usepackage{graphicx}
\usepackage{dsfont}
\usepackage{cite}
\usepackage{amsmath}
\usepackage[T1]{fontenc}
\newcommand\pubnumber{MCNET-16-45}
\newcommand\pubdate{\today}

\newcommand{\FDF}{\left(\varphi^\dagger\overleftrightarrow{D}_\mu\varphi\right)}
\newcommand{\FDFI}{\left(\varphi^\dagger\overleftrightarrow{D}^I_\mu \varphi\right)}

\def\institute{Centre for Cosmology, Particle Physics and Phenomenology (CP3),\\
Universit\'e catholique de Louvain, B-1348 Louvain-la-Neuve, Belgium\\
and\\
Nikhef Theory Group, Science Park 105, 1098 XG Amsterdam, The Netherlands}
\def\support{\footnote{Work supported by the European Union as part of the FP7 Marie Curie
Initial Training Network MCnetITN (PITN-GA-2012-315877).}}

\def\Title#1{\begin{center} {\Large #1 } \end{center}}
\def\Author#1{\begin{center}{ \sc #1} \end{center}}
\def\Address#1{\begin{center}{ \it #1} \end{center}}

\newcommand\pubblock{\rightline{\begin{tabular}{l} \pubnumber\\
         \pubdate  \end{tabular}}}
\newenvironment{Abstract}{\begin{quotation}  }{\end{quotation}}
\newenvironment{Presented}{\begin{quotation} \begin{center} 
             PRESENTED AT\end{center}\bigskip 
      \begin{center}\begin{large}}{\end{large}\end{center} \end{quotation}}
\def\Acknowledgements{\bigskip  \bigskip \begin{center} \begin{large}
             \bf ACKNOWLEDGEMENTS \end{large}\end{center}}
\input econfmacros.tex
\begin{document}
\begin{titlepage}
\pubblock

\vfill
\Title{Top and Electroweak bosons and Higgs}
\vfill
\Author{Eleni Vryonidou\support}
\Address{\institute}
\vfill
\begin{Abstract}
In this talk I review recent progress in the computation of processes involving top quarks in the framework of Standard Model Effective Theory at NLO in QCD. In particular I discuss the impact of higher-dimensional operators on top pair production in association with a photon, a $Z$ boson and a Higgs. Results are obtained within the automated framework of {\sc MadGraph5\_aMC@NLO}.
\end{Abstract}
\vfill
\begin{Presented}
$9^{th}$ International Workshop on Top Quark Physics\\
Olomouc, Czech Republic,  September 19--23, 2016
\end{Presented}
\vfill
\end{titlepage}
\def\thefootnote{\fnsymbol{footnote}}
\setcounter{footnote}{0}

\section{Introduction}
\vspace{-0.1cm}
With Run-II of the LHC, a plethora of precise top-quark measurements 
is expected. In addition to the main production processes: pair and single top production, results for the associated production of top quarks with a vector boson and a Higgs have been collected by CMS and ATLAS. Remarkable progress has been achieved in providing precise theoretical predictions for this class of processes in the Standard Model (SM).  In particular the $t\bar tH$ process is known at next-to-leading order (NLO) in QCD
\cite{Beenakker:2002nc,Dawson:2003zu,Frederix:2011zi,Garzelli:2011vp},
with off-shell effects \cite{Denner:2015yca}, and at NLO in electroweak (EW)
\cite{Frixione:2015zaa,Hartanto:2015uka}. Results for $t\bar{t}Z$ have been presented at NLO in QCD in \cite{Garzelli:2012bn,Frederix:2011zi} and NLO in EW in \cite{Frixione:2015zaa}, while $t\bar{t}\gamma$ is known at NLO in QCD \cite{Frederix:2011zi,Kardos:2014zba}. A detailed phenomenological study of SM top production in association with a Higgs and electroweak bosons can be found in \cite{Maltoni:2015ena}.

Precise predictions for deviations from the SM will become equally
important at the LHC Run II. The Standard Model Effective Field Theory (SMEFT) provides a powerful
framework to consistently and systematically
describe deviations from the SM via higher-dimension operators which modify the SM Lagrangian as follows: 
\begin{equation}  \mathcal{L}= \mathcal{L}_{SM}+\sum_i \frac{C_i}{\Lambda^2} O_i + \mathcal{O}(\Lambda^{-4}).
\end{equation}   
Recently fully differential NLO QCD corrections to top-quark processes in the EFT have started to become
available. These include the top-decay processes \cite{Zhang:2013xya,Zhang:2014rja}, 
single-top production triggered by flavor-changing neutral interactions \cite{Degrande:2014tta},
 top-quark pair production and single top production \cite{Franzosi:2015osa,Zhang:2016omx}. 
QCD corrections are found to have a large and nontrivial impact on both the total cross sections and the differential distributions. NLO results come with reduced theoretical uncertainties and can therefore play an important role in extracting more
 reliable information in the context of global EFT fits \cite{Buckley:2015nca,Buckley:2015lku}. 

In this talk I focus on the recent computation of $t\bar{t}Z$,  $t\bar{t}\gamma$ and $t\bar{t}H$ at dimension-six at NLO in QCD discussed in more detail in \cite{Bylund:2016phk,Maltoni:2016yxb}. 
These results have become available within the {\sc MadGraph5\_aMC@NLO} framework \cite{Alwall:2014hca}.  {\sc FeynRules} and {\sc NLOCT} are used to obtain a UFO model \cite{Alloul:2013bka, Degrande:2014vpa,Degrande:2011ua,deAquino:2011ub} which is then imported into {\sc MadGraph5\_aMC@NLO} to provide NLO accurate results. 

\vspace{-0.3cm}
\section{$t\bar{t}Z/\gamma$ in the EFT}
The operators contributing to $t\bar{t}Z/\gamma$ production up to dimension-six are the following \cite{AguilarSaavedra:2008zc,Grzadkowski:2010es}:
\begin{eqnarray}\nonumber
	&O_{\varphi Q}^{(3)} =i\frac{1}{2}y_t^2 \FDFI (\bar{Q}\gamma^\mu\tau^I
	Q) \label{eq:Ofq3}, \,\,\, O_{\varphi Q}^{(1)} =i\frac{1}{2}y_t^2 \FDF
	(\bar{Q}\gamma^\mu Q) \\ \nonumber &O_{\varphi t} =i\frac{1}{2}y_t^2 \FDF
	(\bar{t}\gamma^\mu t), \,\,\, O_{tW}=y_tg_w(\bar{Q}\sigma^{\mu\nu}\tau^It)\tilde{\varphi}W_{\mu\nu}^I
	\\ &O_{tB}=y_tg_Y(\bar{Q}\sigma^{\mu\nu}t)\tilde{\varphi}B_{\mu\nu}\,\,\, \textrm{and}\,\, O_{tG}=y_tg_s(\bar{Q}\sigma^{\mu\nu}T^At)\tilde{\varphi}G_{\mu\nu}^A\,.
	\label{eq:Otf} 
\end{eqnarray} 

\begin{table}[h]
\renewcommand{\arraystretch}{1.4}
\small
\begin{center}
\begin{tabular}{cccccc}
\hline
   13TeV &  $\mathcal{O}_{tG}$ & $\mathcal{O}^{(3)}_{\phi Q}$ &  $\mathcal{O}_{\phi t}$ &  $\mathcal{O}_{tW}$ \\
\hline
 $\sigma_{i,LO}^{(1)}$  & $ 286.7^{ +38.2 \% }_{ -25.5 \% }$ & $ 78.3^{ +40.4 \% }_{ -26.6 \% }  $ & $ 51.6^{ +40.1 \% }_{ -26.4 \% }$ & $ -0.20^{ +88.0 \% }_{ -230.0 \% }$  \\
  $\sigma_{i,NLO}^{(1)}$ & $ 310.5^{ +5.4 \% }_{ -9.7 \% }$   & $ 90.6^{ +7.1 \% }_{ -11.0 \% } $ & $ 57.5^{ +5.8 \% }_{ -10.3 \% } $ & $ -1.7^{ +31.3 \% }_{ -49.1 \% }$  \\ 
 $K$-factor  & 1.08 & 1.16 & 1.11 & 8.5\\
 $\sigma_{ii,LO}^{(2)}$  & $ 258.5^{ +49.7 \% }_{ -30.4 \% } $  & $ 2.8^{ +39.7 \% }_{ -26.9 \% } $ & $ 2.9^{ +39.7 \% }_{ -26.7 \% } $ & $ 20.9^{ +44.3 \% }_{ -28.3 \% } $ \\
 $\sigma_{ii,NLO}^{(2)}$  & $ 244.5^{ +4.2 \% }_{ -8.1 \% } $ & $ 3.8^{ +13.2 \% }_{ -14.4 \% } $ & $ 3.9^{ +13.8 \% }_{ -14.6 \% } $ & $ 24.2^{ +6.2 \% }_{ -11.2 \% } $  \\
  $K$-factor  & 0.95 & 1.4 & 1.3 & 1.2\\
\hline
\end{tabular}
\end{center}
 \caption{\label{tab:sigmattZ13} Cross sections (in fb) and corresponding $K$-factors for $t\bar{t}Z$ production at the LHC at $\sqrt{s} =  13$~TeV for the different dimension-six operators. Percentages correspond to scale uncertainties. }  
\end{table}
The above operators form a complete set that parameterises
the top-quark interactions to the gluon and the electroweak gauge bosons,
which contribute at $\mathcal{O}(\Lambda^{-2})$. The contributions of these operators to $t\bar{t}Z$, $t\bar{t}\gamma$ and $t\bar{t}\mu^+ \mu^-$ are computed in \cite{Bylund:2016phk} at NLO in QCD. 
As an example we show results for $t\bar{t}Z$ in Table \ref{tab:sigmattZ13}, where the following notation is used:
\begin{equation} 
	\sigma=\sigma_{SM}+\sum_i
	\frac{C_i}{(\Lambda/1\textrm{TeV})^2}\sigma_i^{(1)}+\sum_{i\le j}
	\frac{C_i C_j}{(\Lambda/1\textrm{TeV})^4}\sigma_{ij}^{(2)}\,.
	\label{sigma}
\end{equation} 
\begin{figure}[h]
\begin{minipage}[h]{0.5\linewidth}
\centering
\includegraphics[trim=1cm 0 0 0, scale =1.03]{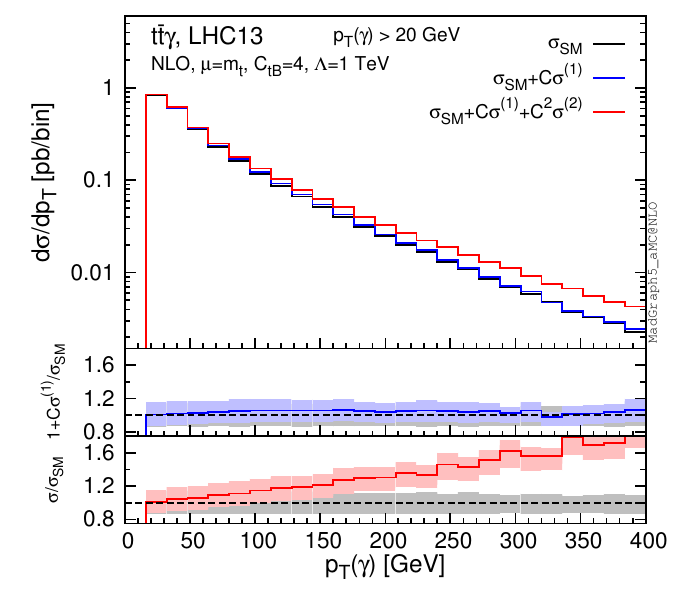}
\end{minipage}
\hspace{0.5cm}
\begin{minipage}[h]{0.5\linewidth}
\centering
\includegraphics[trim=4cm 0  2cm 0,scale=0.9]{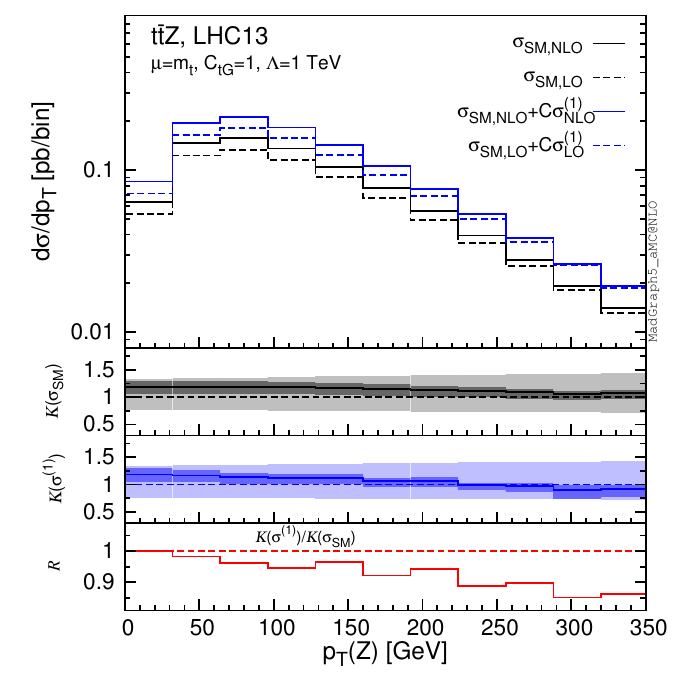}
\end{minipage}
\caption{\label{fig:Kfactors_13} Left: $p_T$ of the photon in $t\bar{t}\gamma$ at 13 TeV for $C_{tB}=4, \Lambda=1$~TeV and ratio over the SM. Right: $p_T$ of the $Z$ in $t\bar{t}Z$ for $C_{tG}=1, \Lambda=1$~TeV. Comparison between the SM and the interference term differential $K$-factors.} 
\end{figure}

Differential distributions for $t\bar{t}Z$, $t\bar{t}\gamma$ and $t\bar{t}\mu^+ \mu^-$ are also obtained in \cite{Bylund:2016phk}. We show some representative results in Fig.~\ref{fig:Kfactors_13} for the $Z$ and photon $p_T$, which demonstrate the difference between the SM and dimension-6 shapes and $K$-factors. Results for all processes are summarised in Fig.~\ref{fig:ttz} which shows the contribution of the various operators to the different processes at LO and NLO.  We note here that the same set of operators impacts the gluon fusion contribution to $HZ$ production and $e^+e^-\to t \bar{t}$, for which results are also presented in \cite{Bylund:2016phk}.

\renewcommand{\arraystretch}{1.8}
\newcommand{\xs}[7]{$#1^{+#2+#6+#4}_{-#3-#7-#5}$}
 \begin{figure}[h]
 \begin{minipage}[t]{0.5\linewidth}
 \vspace{0pt}
 \centering
\includegraphics[width=\linewidth]{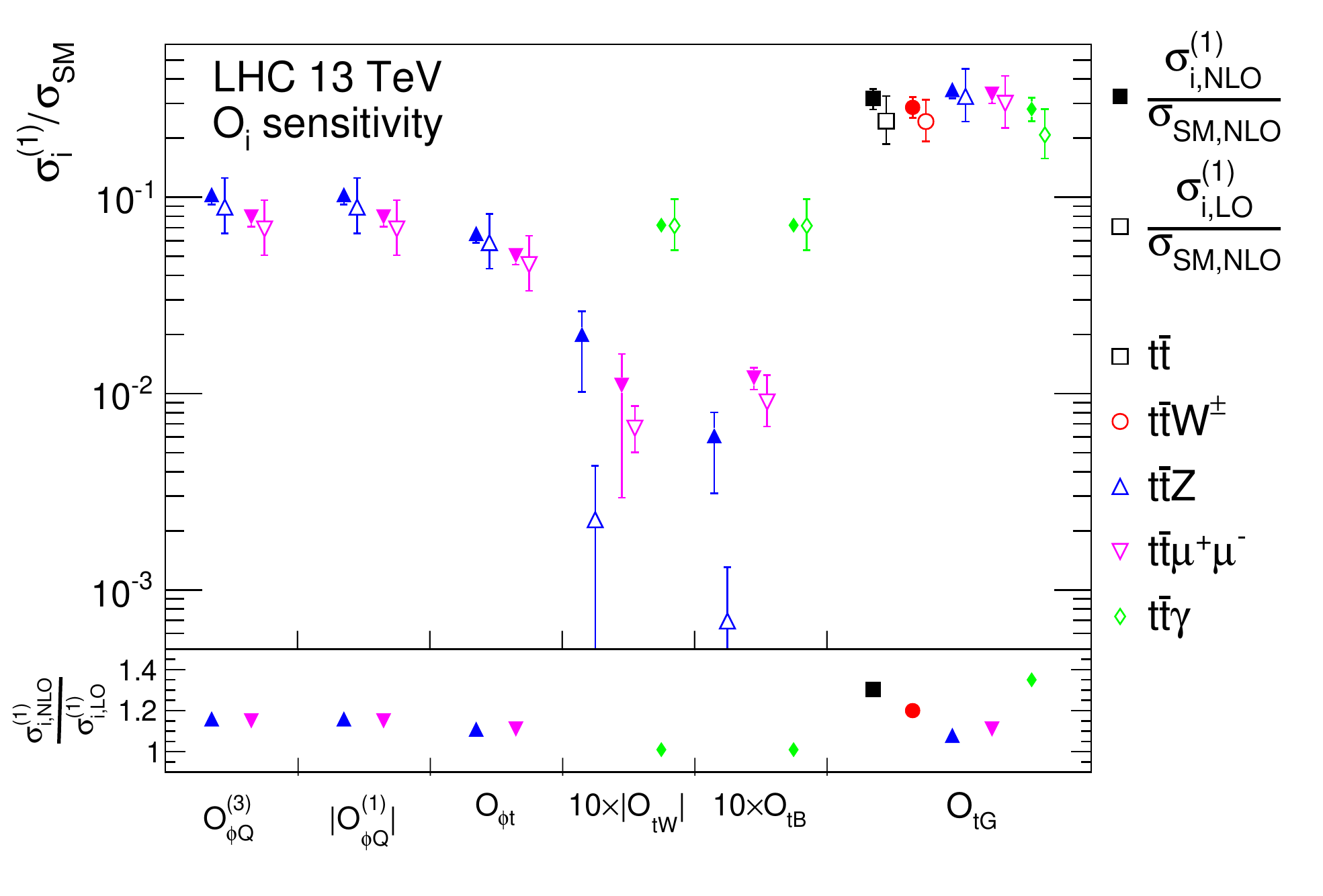}
\caption{Sensitivity of various top quark processes to the various operators
	shown at LO and NLO at 13 TeV. $K$-factors are also shown for
	$\sigma^{(1)}_i$ as well as the scale uncertainties. 
	\label{fig:ttz}}
 \end{minipage}	
  \hspace{0.5cm}
 \begin{minipage}[t]{0.5\linewidth}
 \vspace{0pt}
 \tiny
\begin{tabular}{llllll}
		\hline
		13 TeV &$\sigma$ NLO &$K$-factors
		\\\hline
$\sigma_{SM}$& \xs{0.507}{0.030}{0.048}{0.007}{0.008}{0.000}{0.000}&1.09\\
$\sigma_{t\phi}$& \xs{-0.062}{0.006}{0.004}{0.001}{0.001}{0.001}{0.001}&1.13\\
$\sigma_{\phi G}$& \xs{0.872}{0.131}{0.123}{0.013}{0.016}{0.037}{0.035}&1.39\\
$\sigma_{tG}$&\xs{0.503}{0.025}{0.046}{0.007}{0.008}{0.001}{0.003}&1.07\\
$\sigma_{t\phi,t\phi} $&\xs{0.0019}{0.0001}{0.0002}{0.000}{0.000}{0.0001}{0.000}&1.17\\
$\sigma_{\phi G,\phi G} $&\xs{1.021}{0.204}{0.178}{0.024}{0.029}{0.096}{0.085}&1.58\\
$\sigma_{tG,tG}$& \xs{0.674}{0.036}{0.067}{0.016}{0.019}{0.004}{0.007}&1.04\\
$\sigma_{t\phi,\phi G}$ &\xs{-0.053}{0.008}{0.008}{0.001}{0.001}{0.003}{0.004}&1.42\\
$\sigma_{t\phi,tG}$& \xs{-0.031}{0.003}{0.002}{0.000}{0.000}{0.000}{0.000}&1.10\\
$\sigma_{\phi G,tG}$& \xs{0.859}{0.127}{0.126}{0.017}{0.022}{0.021}{0.020}&1.37\\
\hline
\label{table:tth}
\end{tabular}
\caption{NLO cross sections in pb for $ttH$ at 13 TeV and corresponding $K$-factors. Scale, EFT scale and 
PDF uncertainties are also included.}
\end{minipage}
\end{figure}

\section{$t\bar{t}H$ in the EFT}
For $t\bar{t}H$ production we consider the following operators \cite{Maltoni:2016yxb}:
\begin{eqnarray}\nonumber 
	&O_{t\phi} = y_t^3 \left( \phi^\dagger\phi \right)\left( \bar Qt \right)
	\tilde\phi \,,\,\,O_{\phi G} = y_t^2 \left( \phi^\dagger\phi \right) G_{\mu\nu}^A G^{A\mu\nu}\,,
        \\
	&O_{tG} = y_t g_s (\bar Q\sigma^{\mu\nu}T^A t)\tilde\phi G_{\mu\nu}^A\,.
	\label{operators}
\end{eqnarray}
The three operators mix under RG flow, ~$O_{tG}$ mixes into $O_{\phi G}$, and both of them
mix into $O_{t \phi}$. Results for the cross sections at 13 TeV are summarised in Table \ref{table:tth} using the notation of Eq. \eqref{sigma}. The renormalisation and factorisation scale, 
EFT scale and PDF uncertainties are also shown. The EFT scale uncertainty is associated with the missing higher order corrections to the operators. It is obtained by computing the EFT cross section at a different EFT scale and then evolving this result taking mixing and running effects into account. A detailed discussion of RG effects and a comparison between full NLO  and RG corrections is presented in \cite{Maltoni:2016yxb}, demonstrating that RG corrections are not a good approximation of the full NLO result.
\begin{figure}[h]
 \begin{minipage}[t]{0.5\linewidth}
\centering
\includegraphics[width=.99\linewidth]{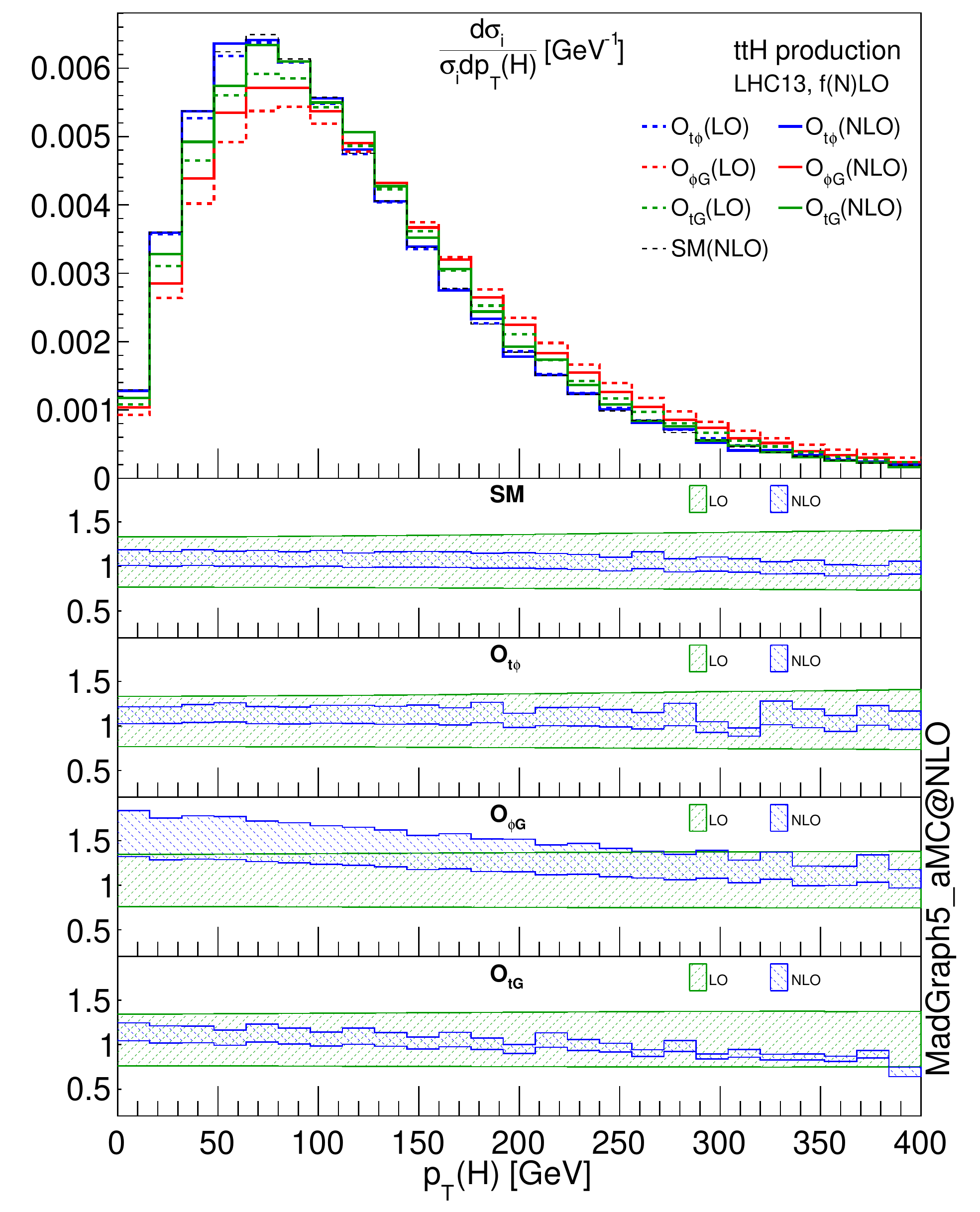}
\end{minipage}
\hspace{0.5cm}
 \begin{minipage}[t]{0.5\linewidth}
 \centering
 \includegraphics[width=.99\linewidth]{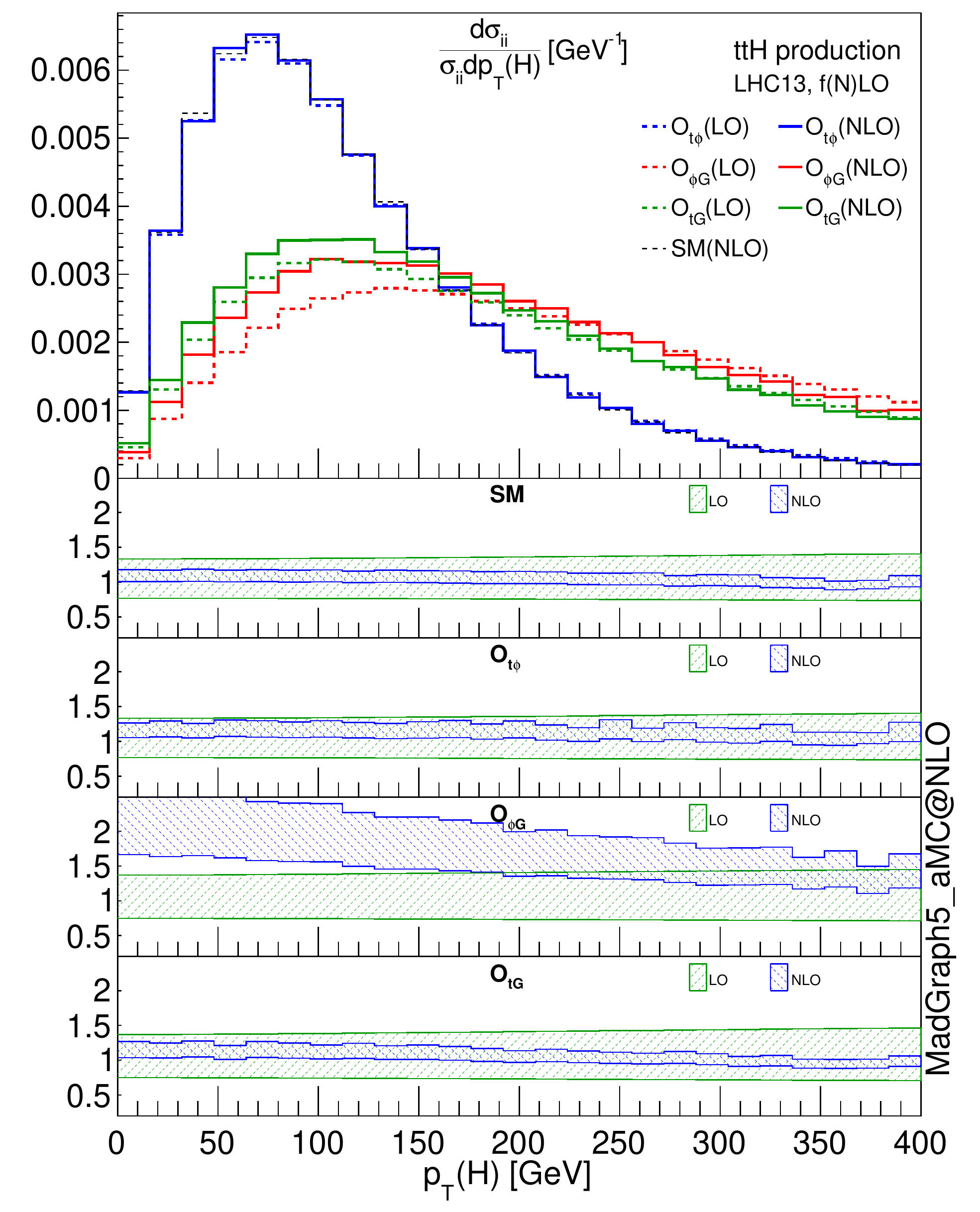}
 \end{minipage}
\caption{\label{fig:pt1} 
Transverse momentum distributions of the Higgs boson in $t\bar{t}H$ at 13 TeV, normalised.  Left: interference contributions from $\sigma_i$. Right:
squared contributions $\sigma_{ii}$.  SM contributions and individual operator
contributions are displayed.  Lower panels give the $K$ factors and $\mu_{R,F}$ uncertainties.} 
\end{figure}
 
Differential distributions can be obtained both at fixed-order and matched to the parton shower.  Figure \ref{fig:pt1} shows the $p_T$ of the Higgs in $t\bar{t}H$ at 13 TeV at LO and NLO (fixed-order), demonstrating different shapes between different operators and non-flat $K$-factors. 

\begin{figure}[h]
		\begin{center}
			\includegraphics[width=.48\linewidth]{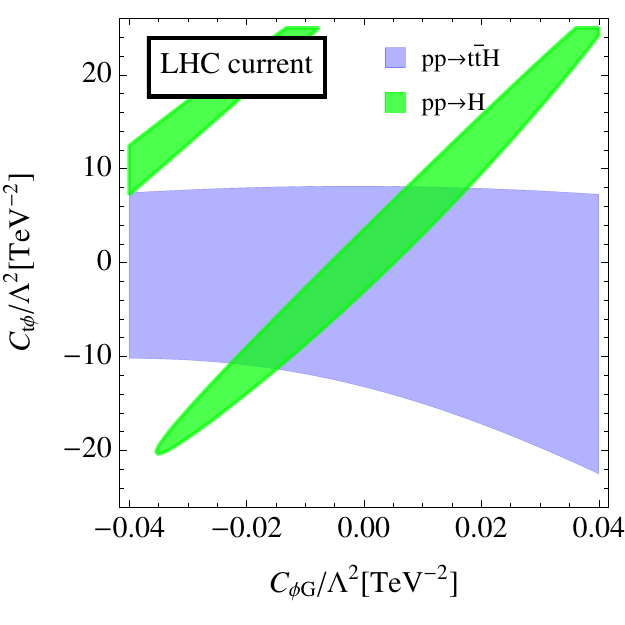}
			\includegraphics[width=.48\linewidth]{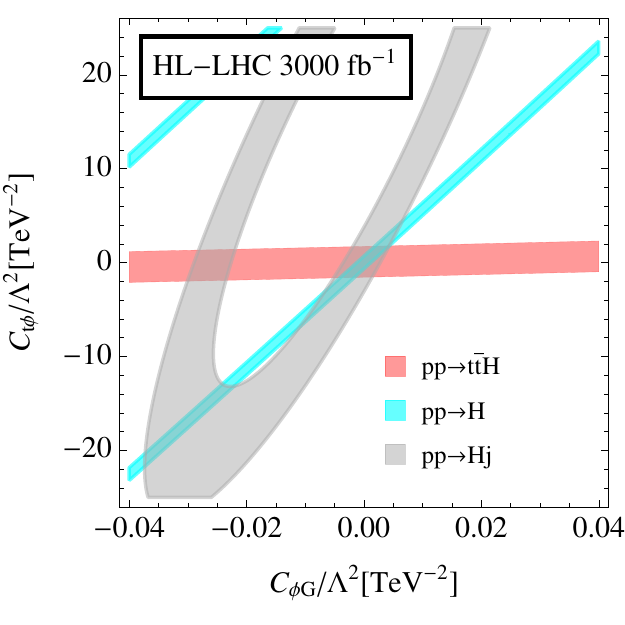}
		\end{center}
		\caption{Allowed region in $O_{t \phi}$-$O_{\phi G}$ plane at 95\% confidence level.
		Left: current constraints.  Right: future projection at HL-LHC.}
		\label{fig:tffg}
	\end{figure}

Finally a connection 
between the top and Higgs sectors is drawn by also considering the loop-induced processes $gg\to H$, $p p \to H j $ and $g g \to H H$ for which 
predictions at dimension-6 are given for the operators of Eq.~\eqref{operators}. The contributions of the chromomagnetic operator ($O_{t G}$) to $Hj$ and $HH$ are obtained for the first time. Current LHC results for single Higgs and $t\bar{t}H$ production  and projections for the High Luminocity (HL) LHC can be used to extract current and potential limits
on the operator coefficients. An example of a two operator fit is shown in Fig. \ref{fig:tffg}, where the degeneracy between $O_{t \phi}$-$O_{\phi G}$ is broken by considering both single Higgs and $t\bar{t}H$ results. A more detailed discussion is given in \cite{Maltoni:2016yxb}.

\section{Summary}
In these proceedings I summarised the computation of top quark processes, in particular $t\bar{t}V$ and $t\bar{t}H$ in the SMEFT at NLO in QCD within the automated {\sc MG5\_aMC} framework. Results for loop-induced processes can also be extracted in the same setup. These results can be readily used to improve the results of global EFT fits in the top sector. 
\Acknowledgements
I would like to thank O. B. Bylund, F. Maltoni, I. Tsinikos, and C. Zhang for their collaboration on this work.

\end{document}

%% file: econfmacros.tex
\def\beq{\begin{equation}}
\def\eeq#1{\label{#1}\end{equation}}
\def\eeqn{\end{equation}}

\def\beqa{\begin{eqnarray}}
\def\eeqa#1{\label{#1}\end{eqnarray}}
\def\eeqan{\end{eqnarray}}

\let\bar=\overbar

\def\Dslash{\not{\hbox{\kern-4pt $D$}}}
\def\dslash{\not{\hbox{\kern-2pt $\del$}}}

\def\msb{{\bar{\ssstyle M \kern -1pt S}}}

%% file: eprint.bbl
\begin{thebibliography}{99}

%%
%%  bibliographic items can be constructed using the LaTeX format in SPIRES:
%%    see    http://www.slac.stanford.edu/spires/hep/latex.html
%%  SPIRES will also supply the CITATION line information; please include it.
%%

%\cite{Beenakker:2002nc}
\bibitem{Beenakker:2002nc} 
  W.~Beenakker, S.~Dittmaier, M.~Kramer, B.~Plumper, M.~Spira and P.~M.~Zerwas,
  %``NLO QCD corrections to t anti-t H production in hadron collisions,''
  Nucl.\ Phys.\ B {\bf 653}, 151 (2003)
  doi:10.1016/S0550-3213(03)00044-0
  [hep-ph/0211352].
  %%CITATION = doi:10.1016/S0550-3213(03)00044-0;%%
  %367 citations counted in INSPIRE as of 28 Nov 2016


%\cite{Dawson:2003zu}
\bibitem{Dawson:2003zu} 
  S.~Dawson, C.~Jackson, L.~H.~Orr, L.~Reina and D.~Wackeroth,
  %``Associated Higgs production with top quarks at the large hadron collider: NLO QCD corrections,''
  Phys.\ Rev.\ D {\bf 68}, 034022 (2003)
  doi:10.1103/PhysRevD.68.034022
  [hep-ph/0305087].
  %%CITATION = doi:10.1103/PhysRevD.68.034022;%%
  %235 citations counted in INSPIRE as of 28 Nov 2016


%\cite{Frederix:2011zi}
\bibitem{Frederix:2011zi} 
  R.~Frederix, S.~Frixione, V.~Hirschi, F.~Maltoni, R.~Pittau and P.~Torrielli,
  %``Scalar and pseudoscalar Higgs production in association with a top?antitop pair,''
  Phys.\ Lett.\ B {\bf 701}, 427 (2011)
  doi:10.1016/j.physletb.2011.06.012
  [arXiv:1104.5613 [hep-ph]].
  %%CITATION = doi:10.1016/j.physletb.2011.06.012;%%
  %123 citations counted in INSPIRE as of 28 Nov 2016


%\cite{Garzelli:2011vp}
\bibitem{Garzelli:2011vp} 
  M.~V.~Garzelli, A.~Kardos, C.~G.~Papadopoulos and Z.~Trocsanyi,
  %``Standard Model Higgs boson production in association with a top anti-top pair at NLO with parton showering,''
  Europhys.\ Lett.\  {\bf 96}, 11001 (2011)
  doi:10.1209/0295-5075/96/11001
  [arXiv:1108.0387 [hep-ph]].
  %%CITATION = doi:10.1209/0295-5075/96/11001;%%
  %90 citations counted in INSPIRE as of 28 Nov 2016


%\cite{Denner:2015yca}
\bibitem{Denner:2015yca} 
  A.~Denner and R.~Feger,
  %``NLO QCD corrections to off-shell top-antitop production with leptonic decays in association with a Higgs boson at the LHC,''
  JHEP {\bf 1511}, 209 (2015)
  doi:10.1007/JHEP11(2015)209
  [arXiv:1506.07448 [hep-ph]].
  %%CITATION = doi:10.1007/JHEP11(2015)209;%%
  %34 citations counted in INSPIRE as of 28 Nov 2016


%\cite{Frixione:2015zaa}
\bibitem{Frixione:2015zaa} 
  S.~Frixione, V.~Hirschi, D.~Pagani, H.-S.~Shao and M.~Zaro,
  %``Electroweak and QCD corrections to top-pair hadroproduction in association with heavy bosons,''
  JHEP {\bf 1506}, 184 (2015)
  doi:10.1007/JHEP06(2015)184
  [arXiv:1504.03446 [hep-ph]].
  %%CITATION = doi:10.1007/JHEP06(2015)184;%%
  %49 citations counted in INSPIRE as of 28 Nov 2016


%\cite{Hartanto:2015uka}
\bibitem{Hartanto:2015uka} 
  H.~B.~Hartanto, B.~Jager, L.~Reina and D.~Wackeroth,
  %``Higgs boson production in association with top quarks in the POWHEG BOX,''
  Phys.\ Rev.\ D {\bf 91}, no. 9, 094003 (2015)
  doi:10.1103/PhysRevD.91.094003
  [arXiv:1501.04498 [hep-ph]].
  %%CITATION = doi:10.1103/PhysRevD.91.094003;%%
  %14 citations counted in INSPIRE as of 28 Nov 2016


%\cite{Garzelli:2012bn}
\bibitem{Garzelli:2012bn} 
  M.~V.~Garzelli, A.~Kardos, C.~G.~Papadopoulos and Z.~Trocsanyi,
  %``t $\bar{t}$ $W^{+-}$ and t $\bar{t}$ Z Hadroproduction at NLO accuracy in QCD with Parton Shower and Hadronization effects,''
  JHEP {\bf 1211}, 056 (2012)
  doi:10.1007/JHEP11(2012)056
  [arXiv:1208.2665 [hep-ph]].
  %%CITATION = doi:10.1007/JHEP11(2012)056;%%
  %174 citations counted in INSPIRE as of 28 Nov 2016


%\cite{Kardos:2014zba}
\bibitem{Kardos:2014zba} 
  A.~Kardos and Z.~Trócsányi,
  %``Hadroproduction of t anti-t pair in association with an isolated photon at NLO accuracy matched with parton shower,''
  JHEP {\bf 1505}, 090 (2015)
  doi:10.1007/JHEP05(2015)090
  [arXiv:1406.2324 [hep-ph]].
  %%CITATION = doi:10.1007/JHEP05(2015)090;%%
  %7 citations counted in INSPIRE as of 28 Nov 2016


%\cite{Maltoni:2015ena}
\bibitem{Maltoni:2015ena} 
  F.~Maltoni, D.~Pagani and I.~Tsinikos,
  %``Associated production of a top-quark pair with vector bosons at NLO in QCD: impact on $ \mathrm{t}\overline{\mathrm{t}}\mathrm{H} $ searches at the LHC,''
  JHEP {\bf 1602}, 113 (2016)
  doi:10.1007/JHEP02(2016)113
  [arXiv:1507.05640 [hep-ph]].
  %%CITATION = doi:10.1007/JHEP02(2016)113;%%
  %23 citations counted in INSPIRE as of 28 Nov 2016


%\cite{Zhang:2013xya}
\bibitem{Zhang:2013xya} 
  C.~Zhang and F.~Maltoni,
  %``Top-quark decay into Higgs boson and a light quark at next-to-leading order in QCD,''
  Phys.\ Rev.\ D {\bf 88}, 054005 (2013)
  doi:10.1103/PhysRevD.88.054005
  [arXiv:1305.7386 [hep-ph]].
  %%CITATION = doi:10.1103/PhysRevD.88.054005;%%
  %33 citations counted in INSPIRE as of 28 Nov 2016


%\cite{Zhang:2014rja}
\bibitem{Zhang:2014rja} 
  C.~Zhang,
  %``Effective field theory approach to top-quark decay at next-to-leading order in QCD,''
  Phys.\ Rev.\ D {\bf 90}, no. 1, 014008 (2014)
  doi:10.1103/PhysRevD.90.014008
  [arXiv:1404.1264 [hep-ph]].
  %%CITATION = doi:10.1103/PhysRevD.90.014008;%%
  %23 citations counted in INSPIRE as of 28 Nov 2016


%\cite{Degrande:2014tta}
\bibitem{Degrande:2014tta} 
  C.~Degrande, F.~Maltoni, J.~Wang and C.~Zhang,
  %``Automatic computations at next-to-leading order in QCD for top-quark flavor-changing neutral processes,''
  Phys.\ Rev.\ D {\bf 91}, 034024 (2015)
  doi:10.1103/PhysRevD.91.034024
  [arXiv:1412.5594 [hep-ph]].
  %%CITATION = doi:10.1103/PhysRevD.91.034024;%%
  %11 citations counted in INSPIRE as of 28 Nov 2016


%\cite{Franzosi:2015osa}
\bibitem{Franzosi:2015osa} 
  D.~Buarque Franzosi and C.~Zhang,
  %``Probing the top-quark chromomagnetic dipole moment at next-to-leading order in QCD,''
  Phys.\ Rev.\ D {\bf 91}, no. 11, 114010 (2015)
  doi:10.1103/PhysRevD.91.114010
  [arXiv:1503.08841 [hep-ph]].
  %%CITATION = doi:10.1103/PhysRevD.91.114010;%%
  %16 citations counted in INSPIRE as of 28 Nov 2016


%\cite{Zhang:2016omx}
\bibitem{Zhang:2016omx} 
  C.~Zhang,
  %``Single Top Production at Next-to-Leading Order in the Standard Model Effective Field Theory,''
  Phys.\ Rev.\ Lett.\  {\bf 116}, no. 16, 162002 (2016)
  doi:10.1103/PhysRevLett.116.162002
  [arXiv:1601.06163 [hep-ph]].
  %%CITATION = doi:10.1103/PhysRevLett.116.162002;%%
  %9 citations counted in INSPIRE as of 28 Nov 2016


%\cite{Buckley:2015nca}
\bibitem{Buckley:2015nca} 
  A.~Buckley, C.~Englert, J.~Ferrando, D.~J.~Miller, L.~Moore, M.~Russell and C.~D.~White,
  %``Global fit of top quark effective theory to data,''
  Phys.\ Rev.\ D {\bf 92}, no. 9, 091501 (2015)
  doi:10.1103/PhysRevD.92.091501
  [arXiv:1506.08845 [hep-ph]].
  %%CITATION = doi:10.1103/PhysRevD.92.091501;%%
  %23 citations counted in INSPIRE as of 28 Nov 2016


%\cite{Buckley:2015lku}
\bibitem{Buckley:2015lku} 
  A.~Buckley, C.~Englert, J.~Ferrando, D.~J.~Miller, L.~Moore, M.~Russell and C.~D.~White,
  %``Constraining top quark effective theory in the LHC Run II era,''
  JHEP {\bf 1604}, 015 (2016)
  doi:10.1007/JHEP04(2016)015
  [arXiv:1512.03360 [hep-ph]].
  %%CITATION = doi:10.1007/JHEP04(2016)015;%%
  %19 citations counted in INSPIRE as of 28 Nov 2016


%\cite{Alwall:2014hca}
\bibitem{Alwall:2014hca} 
  J.~Alwall {\it et al.},
  %``The automated computation of tree-level and next-to-leading order differential cross sections, and their matching to parton shower simulations,''
  JHEP {\bf 1407}, 079 (2014)
  doi:10.1007/JHEP07(2014)079
  [arXiv:1405.0301 [hep-ph]].
  %%CITATION = doi:10.1007/JHEP07(2014)079;%%
  %1436 citations counted in INSPIRE as of 28 Nov 2016


%\cite{Alloul:2013bka}
\bibitem{Alloul:2013bka} 
  A.~Alloul, N.~D.~Christensen, C.~Degrande, C.~Duhr and B.~Fuks,
  %``FeynRules  2.0 - A complete toolbox for tree-level phenomenology,''
  Comput.\ Phys.\ Commun.\  {\bf 185}, 2250 (2014)
  doi:10.1016/j.cpc.2014.04.012
  [arXiv:1310.1921 [hep-ph]].
  %%CITATION = doi:10.1016/j.cpc.2014.04.012;%%
  %501 citations counted in INSPIRE as of 28 Nov 2016


%\cite{Degrande:2014vpa}
\bibitem{Degrande:2014vpa} 
  C.~Degrande,
  %``Automatic evaluation of UV and R2 terms for beyond the Standard Model Lagrangians: a proof-of-principle,''
  Comput.\ Phys.\ Commun.\  {\bf 197}, 239 (2015)
  doi:10.1016/j.cpc.2015.08.015
  [arXiv:1406.3030 [hep-ph]].
  %%CITATION = doi:10.1016/j.cpc.2015.08.015;%%
  %51 citations counted in INSPIRE as of 28 Nov 2016


%\cite{Degrande:2011ua}
\bibitem{Degrande:2011ua} 
  C.~Degrande, C.~Duhr, B.~Fuks, D.~Grellscheid, O.~Mattelaer and T.~Reiter,
  %``UFO - The Universal FeynRules Output,''
  Comput.\ Phys.\ Commun.\  {\bf 183}, 1201 (2012)
  doi:10.1016/j.cpc.2012.01.022
  [arXiv:1108.2040 [hep-ph]].
  %%CITATION = doi:10.1016/j.cpc.2012.01.022;%%
  %349 citations counted in INSPIRE as of 28 Nov 2016


%\cite{deAquino:2011ub}
\bibitem{deAquino:2011ub} 
  P.~de Aquino, W.~Link, F.~Maltoni, O.~Mattelaer and T.~Stelzer,
  %``ALOHA: Automatic Libraries Of Helicity Amplitudes for Feynman Diagram Computations,''
  Comput.\ Phys.\ Commun.\  {\bf 183}, 2254 (2012)
  doi:10.1016/j.cpc.2012.05.004
  [arXiv:1108.2041 [hep-ph]].
  %%CITATION = doi:10.1016/j.cpc.2012.05.004;%%
  %80 citations counted in INSPIRE as of 28 Nov 2016


%\cite{Bylund:2016phk}
\bibitem{Bylund:2016phk} 
  O.~Bessidskaia Bylund, F.~Maltoni, I.~Tsinikos, E.~Vryonidou and C.~Zhang,
  %``Probing top quark neutral couplings in the Standard Model Effective Field Theory at NLO in QCD,''
  JHEP {\bf 1605}, 052 (2016)
  doi:10.1007/JHEP05(2016)052
  [arXiv:1601.08193 [hep-ph]].
  %%CITATION = doi:10.1007/JHEP05(2016)052;%%
  %22 citations counted in INSPIRE as of 28 Nov 2016


%\cite{Maltoni:2016yxb}
\bibitem{Maltoni:2016yxb} 
  F.~Maltoni, E.~Vryonidou and C.~Zhang,
  %``Higgs production in association with a top-antitop pair in the Standard Model Effective Field Theory at NLO in QCD,''
  JHEP {\bf 1610}, 123 (2016)
  doi:10.1007/JHEP10(2016)123
  [arXiv:1607.05330 [hep-ph]].
  %%CITATION = doi:10.1007/JHEP10(2016)123;%%
  %6 citations counted in INSPIRE as of 28 Nov 2016


%\cite{AguilarSaavedra:2008zc}
\bibitem{AguilarSaavedra:2008zc} 
  J.~A.~Aguilar-Saavedra,
  %``A Minimal set of top anomalous couplings,''
  Nucl.\ Phys.\ B {\bf 812}, 181 (2009)
  doi:10.1016/j.nuclphysb.2008.12.012
  [arXiv:0811.3842 [hep-ph]].
  %%CITATION = doi:10.1016/j.nuclphysb.2008.12.012;%%
  %238 citations counted in INSPIRE as of 28 Nov 2016


%\cite{Grzadkowski:2010es}
\bibitem{Grzadkowski:2010es} 
  B.~Grzadkowski, M.~Iskrzynski, M.~Misiak and J.~Rosiek,
  %``Dimension-Six Terms in the Standard Model Lagrangian,''
  JHEP {\bf 1010}, 085 (2010)
  doi:10.1007/JHEP10(2010)085
  [arXiv:1008.4884 [hep-ph]].
  %%CITATION = doi:10.1007/JHEP10(2010)085;%%
  %481 citations counted in INSPIRE as of 28 Nov 2016
















 
\end{thebibliography}
